\documentclass{aa}  
\usepackage{graphicx}
\usepackage{txfonts}
\begin{document}
\title { Instability of P-waves just below the transition region in a global solar wind simulation}
   \author{R. Grappin
          \inst{1}
          J. L\'eorat
          \inst{1}
          R. Pinto
          \inst{1}
          \and
          Y.-M. Wang\inst{2}
          }
   \institute{Observatoire de Paris, LUTH, CNRS,
              92195 Meudon, France\\
              \email{Roland.Grappin@obspm.fr}
         \and
             NRL\\
             \email{ywang@yucca.nrl.navy.mil}
             }

   \date{Received September 15, 1996; accepted March 16, 1997}

 
  \abstract
   { To progress in the understanding of the solar wind and coronal dynamics, numerical modeling must include the transition region within the simulation domain, and not just as a bottom boundary.
   Published simulations including a transition region within the domain often do not take into account any modeling of the heat sources and sinks which generate the corona; in the rare self-consistent simulations including it, the transition region itself appears to be chaotic in space and time.
}
   {Our aim is to investigate the response to perturbations of the solar atmosphere including transition region and wind, and more specifically how wave propagation is modified by the presence of heat sources and sinks, in the simple 1D, hydrodynamical case, including chromosphere and solar wind.}
   {We integrate the time-dependent hydrodynamic equations of the solar wind with spherical symmetry, including conduction, radiative cooling and a prescribed mechanical heat flux. Once a quasi-stationary wind is established, we study the response of the system to pressure oscillations at the photospheric boundary. We use transparent boundary conditions}
   {We find that wavepackets with high enough amplitude propagating upward from the photosphere implode just below the transition region. This implosion is due to the radiative cooling term generating pressure holes close to the wave crests of the wave, which make the wave collapse. In the case where heat sources and sinks are not present in the equations, the wave remains stable whatever the initial wave amplitude, which is compatible with published work.
   }
   {The instability found here is not an instability of the TR itself: on the contrary, instability ceases when the wavepacket enters the TR where conduction is able to balance the cooling. However, the TR as a whole can be destabilized by such implosions, which should be observable when and where the TR is high enough above the optically thick regions.
}


   \maketitle
%

\section{Introduction}
Among the many choices to be made before beginning a numerical study of the corona-solar wind dynamics, one must decide where to place the bottom boundary; in particular, should we include the dense cold layers below the corona, or not?
Including them will increase the numerical cost (as the scale height of the cold layers is small), and, as well, resolving the chromospheric transition region itself will require to increase the spatial and temporal resolutions there.
So, we are tempted not to include the cold layers. The work by Lionello et al. (2001) and by Endeve et al. (2003) shows that interesting results can be obtained in this way, albeit with numerous difficulties.

However, not including the cold layers means not including completeley the transition region, 
so that we are left with a coronal bottom boundary, where we have to chose some (arbitrary) conditions.
If we are not especially interested by the time-dependent atmosphere, 
then this is workable (see Lionello et al. 2001). 
But if we are, then the response of the corona to perturbations might depend much on these arbitrary boundary conditions. 
This might show up when perturbing the bottom boundary (compare the results obtained by using "line-tied", i.e., full reflexion in Aulanier et al. (2005) and those obtained by using complete transparency in Grappin et al. (2005)). 

The transition region itself might be the source of interesting dynamics still to be understood. Indeed, simulations by Suzuki and Inutsuka (2005) (a global 1.5D time-dependent solar wind study, denoted by SI in the following) and Gudiksen and Nordlund (2005) (3D volume of solar atmosphere), consider sucessfully the issue of self-consistent coronal heating. They both find that the transition region is erratic both in time and space. 

The second question is to specify the energy equation. 
The minimal energy equation includes the adiabatic term with $\gamma=5/3$. 
If we are not interested in the dynamics of the transition region, that is all for the energy equation, and we simply start with a given atmosphere temperature profile, possibly adding extra coupling terms in the energy equation to limit the numerical diffusion of the temperature profile.
This is done in De Pontieu et al. (2005) in their work on spicules, and more or less so also (Velli, personal communication) in Del Zanna et al. (2005) in their work on coronal seismology.

Alternatively, if one includes heat sources and sinks, it is reasonable to expect that the dynamics should be modified, even for small amplitude perturbations. 
There is one single simulation of the corona and solar wind which includes heat sources and sinks self-consistently, the one by SI: it used 14000 grid points, which is clearly not generalizable to 2D or 3D.
Our aim here is double: to devise a reduced version of such a 1D model, with a reduced number of grid points, e.g. 300 points instead of 14000, so allowing future generalization to 2D/3D, and use it to study the specific effects on solar atmosphere dynamics (including transition region) of heat sources and sinks.

In this preliminary work, we solve the hydrodynamics equations, not the MHD equations,
and inject pressure waves at the photospheric level. Since we do not expect that plain pressure waves are able to heat the corona, we use a prescribed mechanical energy flux which is a function of heliocentric distance as e.g. in Wang, 1994: hence our model is not self-consistent, contrary to that of SI.

A basic starting point for the understanding of waves in the solar atmosphere is the work by Fleck and Schmitz (1991) and Kalkofen et al. (1994). They studied the response to base perturbations of an isothermal atmosphere, with only adiabatic terms in the energy equation, and a uniform gravity. Recall that 1D pressure oscillations propagate or not, depending on the frequency being larger or smaller thant the cut-off Lamb frequency:
\begin{equation}
\label{Lamb}
\omega_L = c/(2H) = \gamma g/(2c)
\end{equation}
$c$ being the sound speed, $g$ the gravity, $H$ the pressure scale height. 
When propagative, waves have their energy flux about constant in the limit of high frequencies. When evanescent, all the atmosphere oscillates in phase.
This is the asymptotic stationary state. But before this, interesting transients occur, as has been discovered by the above authors.
Long-living transients with the cut-off frequency dominate the spectrum in a large part of the atmosphere and during a long time, except in the very lower layers. 
Fleck and Schmitz have shown that the phenomenon is about the same, whether a given transition region is present or not. 
This was proposed to be a possible explanation for the observed prevalence of 3-min oscillations in chromospheric lines.
How are these results modified when energy sources and sinks are taken into account?

We continue here this work by adding heat sources and sinks to the energy equation. We will show that the transients with cut-off frequency become unstable when they arrive below the TR, at an altitude where radiative cooling begins to be active. They may implode, leading to catastrophic events.
This work will at the same time demonstrate the feasibility of a time-dependent model of a corona-solar wind starting at the photospheric level with a reduced number of grid points.

The plan is the following. Section 2 describes the equations and the method. Section 3 reports the results. The last section is a discussion.


\section{Equations and method}
We consider a spherically symmetric solar corona, and solve the time-dependent gas equations, which are given by the equation for the radial velocity:
\begin{equation}
\label{velocity}
\partial u/\partial t + u \partial u/\partial r + (1/\rho) \partial P/\partial r + GM/r^2 = 0
\end{equation}
Pressure and temperature equations:
\begin{equation}
\label{pressure}
\partial P/\partial t + u \partial P/\partial r + \gamma P div u = Q
\end{equation}
\begin{equation}
\label{temperature}
\partial T/\partial t + u \partial T/\partial r + (\gamma-1) T div u = Q/\rho
\end{equation}
where $\rho=m_p n$. 
The density n is deduced from the pressure and temperature, using the equation of state:
\begin{equation}
\label{temperature}
P=2nkT
\end{equation}
T being the average temperature of ions and electrons.
 The extra, non-adiabatic source term $Q$ is given by either one of the two following models.\\
(a)the "full model" including heat sources and sinks:
\begin{equation}
\label{temperature}
Q = -(\gamma-1) (div F_m + \rho^2\Lambda(T) + div F_c) + \kappa_1 T"
\end{equation}
where $F_m$, $\Lambda(T)$, and $F_c$ are the (prescribed) mechanical energy flux, the radiative cooling term and the conductive flux,
and $\kappa_1 T"$ provides a filtering of the temperature, with $T"$ being a second order derivative weighted by the local mesh $\Delta r$: $T" = \partial^2T/\partial r^2 (\Delta r/<\Delta r >)^2$, $<\Delta r>$ being the average mesh.\\
(b) the "relaxation model" including as the only non adiabatic term a small smoothing term:
\begin{equation}
\label{temperature}
Q = \kappa_1 (T"-T_0(r)")
\end{equation}
where $T_0(r)$ is a prescribed temperature profile, defining the transition region.
In both the full and the relaxation models, the coefficient $\kappa_1$ is small enough, so that its characteristic time scale is much longer than the time scale of the phenomenon studied.

The prescribed phenomenological mechanical flux which leads to global coronal heating is 
\begin{equation}
\label{Fm}
F_m(r) = \hat e_r F_m^0 (R_s/r)^2  exp((R-R_s)/R_H)
\end{equation}
A typical value for the energy flux is $F_m^0 = 10^5 erg/cm^2/s$, and the characteristic scale for energy dissipation is $R_H=R_s$.
The radiation loss function $\Lambda(T)$ is a rough fit to the one by Athay (1986).
More precisely:
\begin{equation}
\label{Lam}
\Lambda=A 10^{-(log_{10}(T/T_M)^2} f(T)  
\end{equation}
with $T_M=0.2 MK$ gives the maximum cooling temperature.
The factor $f(T)$ is between 0 and 1. It is set to zero if the temperature is smaller than a threshold $T_\star$ or larger than 1MK. Between $T_\star$ and $T_{\star\star}=0.02 MK$, $f(T)$ varies linearly between 0 and 1. The low cut-off temperature is set to:
\begin{equation}
\label{tstar}
T_\star = 0.01 MK
\end{equation}
Finally, the factor A is fixed so as to lead to a peak value $\Lambda_{max}=10^{-22} cgs$ (but see Table 1). 
The conductive Spitzer-Harm flux is
\begin{equation}
\label{Fm}
F_c = -\hat e_r \kappa_0 T^{5/2} \partial T/\partial r
\end{equation}
where $\kappa_0 = 10^{-6} cgs$. This value is in between the conductivity for the proton and electron.
This form being highly demanding in computer resources (and mesh size), we reduce steepness of the resulting flux in two ways. First, we use a linear temperature dependence for conductivity at temperature lower than 0.25 MK and pass progressively to the Spitzer-Harm $5/2$ power law between 0.25 MK and 1MK (see Linker et al. (2001)). Second, we limit the resulting flux by requiring that the associated characteristic time $\tau_\star$ be not smaller than a prescribed value $\tau_{lim}$. The final form of the heat flux is thus:
\begin{equation}
\label{limiteur}
F_c =  -\hat e_r (2/7) \kappa_0 \partial T^{N(T)}\partial r 
(1+\tau_\star/\tau_{lim})^{-1}
\end{equation}
with $\tau_\star = \kappa_0 T^{N(T)-1} (\pi/\Delta r)^2/\rho$ and
$\Delta r$ is the local mesh size. Note that in practice, the limiting factor will only be active in the preliminary phase of the runs (run S1 in Table 1).


We use a non-uniform mesh with 300 points following a logarithmic progression of ratio q=1.025. This corresponds to a minimum mesh size being $\Delta r=10^{-4}R_s = 70$ km at the surface, and a maximum mesh $\Delta r=0.4R_s$ at the outer boundary ($r=15 R_s$).
Our temporal scheme is Runge-Kutta of order 3; the spatial scheme is a compact finite difference scheme of order 6 (Lele, 1992) modified to be able to cope with non-uniform grids (J.-M. Le Saout 2003, Grappin et al. 2005). 
Note however that to compute the temperature gradients which appear in the conductive term, we use a scheme or order two, which is more stable.
Finally, we use 6-th order filtering, also defined in Lele (1992). 
Filtering is used in several places. It is used for the velocity field (roughly once every 10 time steps), and for smoothing different quantities before computing the right-hand sides of the equations. Specifically, the (logarithmic) gradients of pressure, density and temperature are systematically filtered; finally the rhs of the velocity equation is filtered: the latter considerably increases the stability of the schema.
Time-step is automatically adapted to the different terms, using dimensional estimations of the characteritstic times.

The boundary conditions are imposed via the characteristic form of the equations. They are: no incoming perturbation at the inner boundary, and, starting with time t=1, a constant depression at the outer boundary, introduced via the ingoing characteristics. This depression stops as soon as the sonic Mach number is reached, since thereafter no incoming signal can progress into the domain from the exterior.

Units used in the following are the solar radius for distance, MK for temperature, and km/s or m/s for velocity. Time indicated in figures and tables is mostly measured in numerical unit, which is 1h30. The list of runs building the stationary corona and wind is given in Table 1, a liste of runs in which we perturb the atmosphere with waves injected at the base is given in Table 2. 

\section{Results}
\label{section3}

\begin{table}
\label{table1}
\begin{tabular}{cccccccc}
\multicolumn{7}{l}{   }      \\
Run & time & $F_m^0$ & $\Lambda_{max}$ & $\kappa_0$ & $\tau_{lim}$ & N \\
S1 & 0,20 & $4 10^4$ & $10^{-21}$ & $10^{-6}$ & $10^{-6}$ & 300\\
S2 & 20,40 & $9 10^4$ & $10^{-22}$ & $10^{-6}$ & $10^{-7}$ & 300 \\
S3 & 40,60 & $9 10^4$ & $10^{-22}$ & $10^{-6}$ & $10^{-7}$ & 300\\
D1 & 60,70 & $9 10^4$ & $10^{-22}$ & $10^{-6}$ & $10^{-7}$ & 599\\
\end{tabular}
\caption[]{
List of runs leading to stationary corona and wind, with heat and source coefficients. Runs S1, S2, S3, D1 follow each other. S1 and S2 respectively correspond to the two phases mentioned in the text. 
S3 is almost identical to S2 (only the viscous terms differ). D1 starts from an extrapolation of run S3 to a more refined mesh and then relaxes.
Time: beginning and end time of each phase; $F_m^0$: base mechanical flux; $\Lambda_{max}$: maximum of cooling term $\Lambda(T)$; $\kappa_0$: conductivity parameter; 
$\tau_{lim}$: characteristic time defining the attenuation of the conductive flux;
N is the resolution
}
\end{table}

\begin{table}
\label{table2}
\begin{tabular}{cccccccc}
\multicolumn{7}{l}{   }      \\
Runs & energy & $\tau$ & $\tau_0$ & ev/prop & amplitude & N \\
A1 & no (1) & 6 min & 7.6 min & prop & 0.4 m/s & 300 \\
A1a & no (1) & 6 min & 7.6 min & prop & 1.3 m/s & 300 \\
A2 & no (1) & 11 min & 7.6 min & ev & 4 m/s & 300 \\
B4 & yes & 10 min & 6-8 min & ev & 4m/s & 300 \\
B40 & yes & 10 min & 6-8 min & ev & 40m/s & 300 \\
C40 & no(2) & 10 min & 6-8 min &ev & 40m/s & 300Ê\\
C400 & no(2) & 10 min & 6-8 min &ev & 400m/s & 300 \\
D40 & yes & 10 min &6-8 min & ev & 40m/s & 599 \\
D80 & yes & 10 min &6-8 min & ev & 80m/s & 599 \\

\end{tabular}
\caption[]{
List of runs with wave injection. runs A1, A2 start injecting waves at time t=0; other runs start from a relaxed wind flow. Runs Bxx, Cxx start from run S3, runs Dxx start from run D1 (see above Table 1).
Energy: yes means energy equation is present; no(1) means coupling of temperature with step-wise constant profile; no(2) means coupling temperature with the stationary profile resulting from integrating the full energy equation. $\tau$ and $\tau_0$ are resp. the period of the injected wave and the cut-off period of the dense layers. Ev/prop: ev means evanescent in dense layers, prop means propagative in dense layers. Amplitude is the base amplitude of the wave. N is the number of grid points.
}
\end{table}

\subsection{Wave transfer in a relaxation model}
\begin{figure}[ht]
\begin{center}
\includegraphics [width=\linewidth]{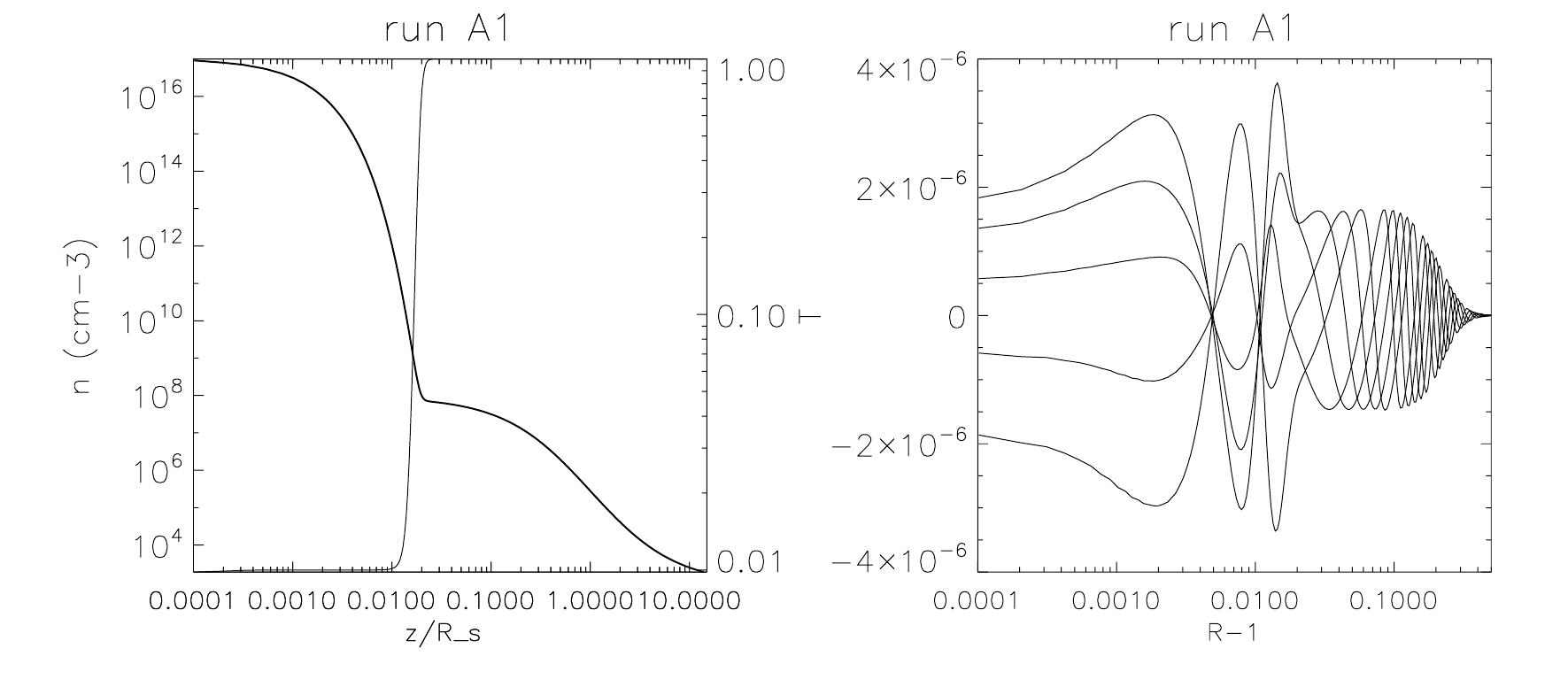}
\caption{Waves in a two-temperature atmosphere (relaxation model):
Left: Density (thick line) and Temperature (plain line) profiles;
Right: run A1, waves with frequency above cut-off, profiles of $(\rho r^2 v_g)^{1/2} u$ in a the (1,1.5) distance range.
}
\label{fig1}
\end{center}
\end{figure}

We first consider in this subsection a two-temperature atmosphere without energy sinks and sources, to check that we recover basic wave properties in a simple configuration.
The atmosphere is made of two parts: a low, dense part at $T=0.01 MK$ and a corona at $T=1 MK$. The transition region lies at about $R=0.015 R_s$.
Apart from that, the parameters are the same as for the full wind model:
domain size $L=14 R_s$, base density is $10^{17} cm^{-3}$, and same grid.
Figure~\ref{fig1} shows the density and temperature profiles (left).
In the right panel, we show the profiles of the (square root of the) energy flux, which result from injecting a monochromatic wave from time t=0 at the base. 
The frequency is $\omega=100$, which is higher than the maximum of cut-off frequency (about $\omega_0 = 90$) in the low-temperature region. 
Hence the quantity plotted should be approximately invariant, except for reflexions,
due possibly to the frequency being too low in the cold region, and most certainly to reflexions at the TR.
The the sudden energy flux decrease after $R=1.1 R_s$ is due to the filtering, which can be made lighter, so that the damping distance is increased.
Filtering should certainly be decreased in view of the future sef-consistent heating of the corona by wave dissipation.

\begin{figure}[ht]
\begin{center}
\includegraphics [width=\linewidth]{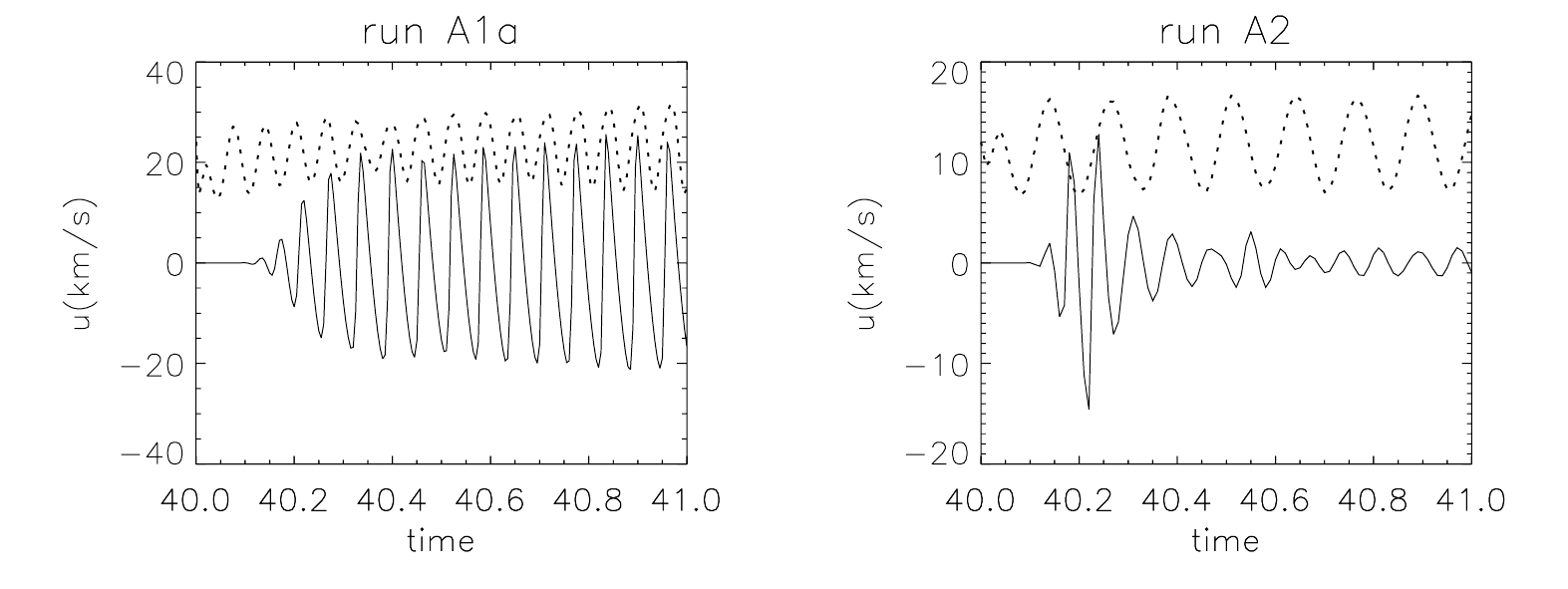}
\caption{Oscillations at TR with two-temperature atmosphere:
velocity at TR versus time (dotted: base velocity);
left: run A1a, propagative incident wave with period 6 min. (below cut-off period = 7.6 min.);
right: run A2, evanescent waves with period 11 min.
}
\label{fig2}
\end{center}
\end{figure}

The main property which interests us here is the very different response of the atmosphere to propagative and evanescent waves explained by Fleck and Schmitz (1991) and Kalkofen et al. (1994), as mentioned in Section 2.
This is shown in Figure~\ref{fig2}. We compare there the velocity response of the TR to waves respectively above and below the cut-off (runs A1a and A2).
The signal at the base is shown as dotted points. 
One sees that, while the oscillation at the TR has the same frequency as injected in the high-frequency case, this is not the case in the evanescent case: at the TR (and in lower layers as well), the frequency becomes very close to the cut-off frequency (period about 7.6 min.).
Note also that the TR response is largest during the first two or three periods.

\subsection{Generation of full wind model}
We give here an account of the method used to generate the wind and hot corona.
We start with an atmosphere uniformly at $6000K$, and a corresponding hydrostatic equilibrium. The initial hydrostatic stratification is too strong to start immediately to integrate the whole set of fluid equations, essentially because the density is very low at the distance of $R=15R_s$, which leads to very small characteristic times.
We thus decide to first raise substantially the temperature atmosphere before integrating the full equations.
This is done during a time limited to $t \le 1$: we integrate only the temperature equation (either with the full energy terms of the plain relaxation term, depending on the model), and the equation of hydrostatic equilibrium to obtain the pressure. We then use the equation of state to deduce the density from pressure and temperature.

\begin{figure}[ht]
\begin{center}
\includegraphics [width=\linewidth]{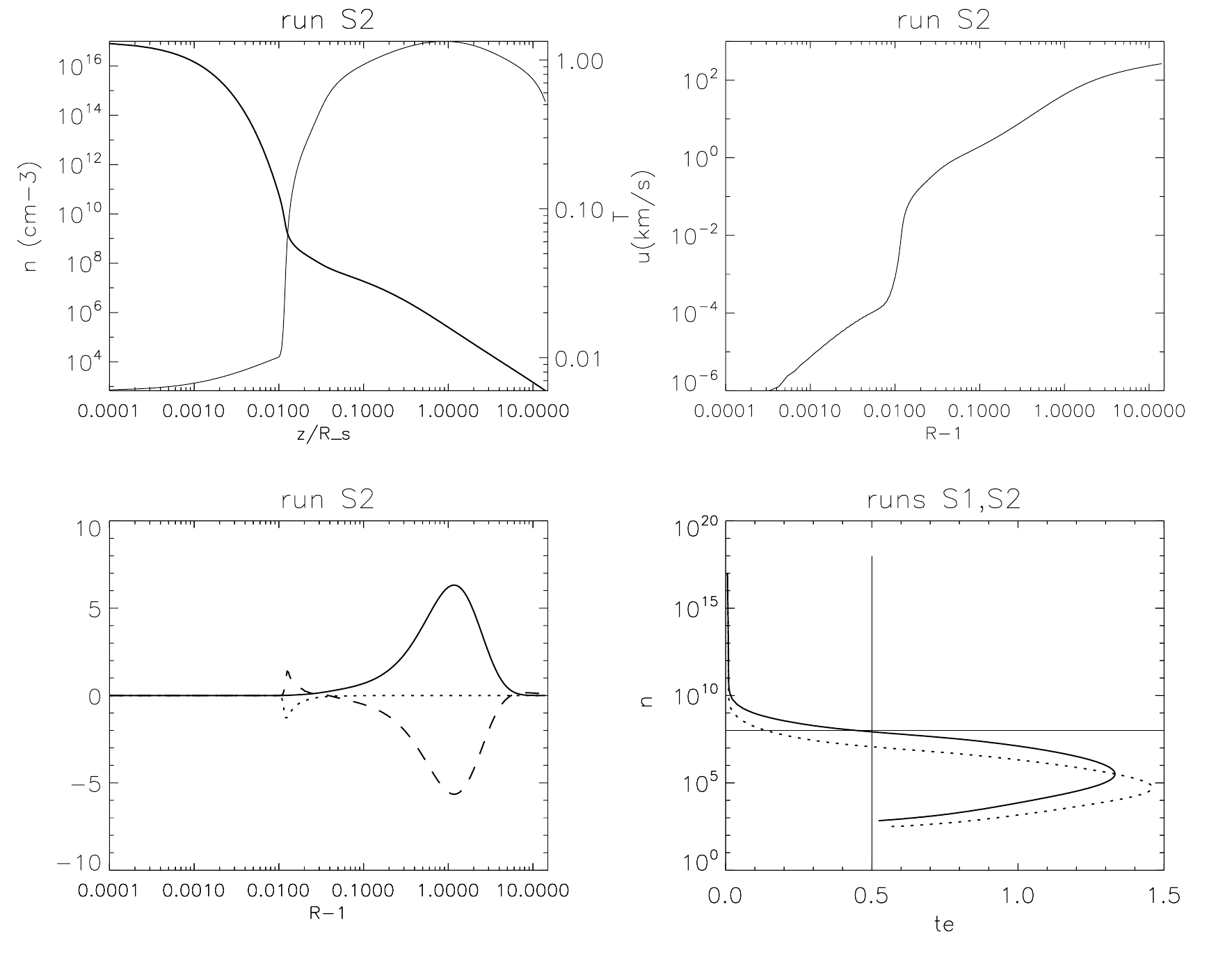}
\caption{
Stationary transonic wind with full heating sources and sinks (end of run S2, see Table 2).
Top left: density and temperature;
top right: velocity;
bottom left: mechanical heating source (thick line), radiative cooling term (dotted),
conductive term (dashed line);
note sources and sinks are as they appear in the temperature equation.
bottom right: density versus temperature, run S2 with thick line, run S1 with dotted line. 
}
\label{fig0}
\end{center}
\end{figure}

After this short hydrostatic heating phase, we integrate the full equations. As stated above, we also perform a depression at the outlet, which forms the transonic wind after some time.
When using the full heat sources and sinks, we use successively two sets of parameters (see runs in Table 1).
The aim is to approach a density close to $n=10^8 cm^{-3}$ where the temperature reaches $T=0.5 MK$. The first set of parameters (run S1) leads to a density a bit below this value (see fourth panel in fig~\ref{fig0}); the second set (all other runs) achieves the desired result.
Note that the first choice of $\tau_{lim}=10^{-6}$ (run S1) limits somewhat the conductive flux, but that the second value (other runs) leads to no limitation at all.
Figure~\ref{fig0} shows several profiles defining the final wind (with set number 2), as well as (bottom right panel) the two temperature/density curves obtained with both sets.
A transition region is seen to occur at about 0.01 $R_s$ = 7 000 km. The temperature has a minimum about 6000 K at the bottom, and a maximum around 1.6 MK, around 4 solar radii. The chromospheric temperature curve is a smooth ramp, leading to a temperature close to 10000 K at the foot of the TR. This ramp is due to the artificial diffusive term included in the temperature and pressure equations (3-4-5).
The sonic Mach number is reached at about 5 solar radii. Note that the heat sources and sinks balance (bottom left panel): conductive heating and radiative loss balance at the TR, while conductive loss and mechanical heating balance (with the help of the adiabatic cooling, not shown) in the corona.

\subsection{Wave transfer in the full wind model}

We now inject waves at the base of the previous stationary atmosphere.
We start from the relaxed wind (run S3) with density about $10^8 cm^{-3}$ where the plasma has temperature $T=0.5MK$ (see figure~\ref{fig0}).

\begin{figure}[ht]
\begin{center}
	\includegraphics [width=\linewidth]{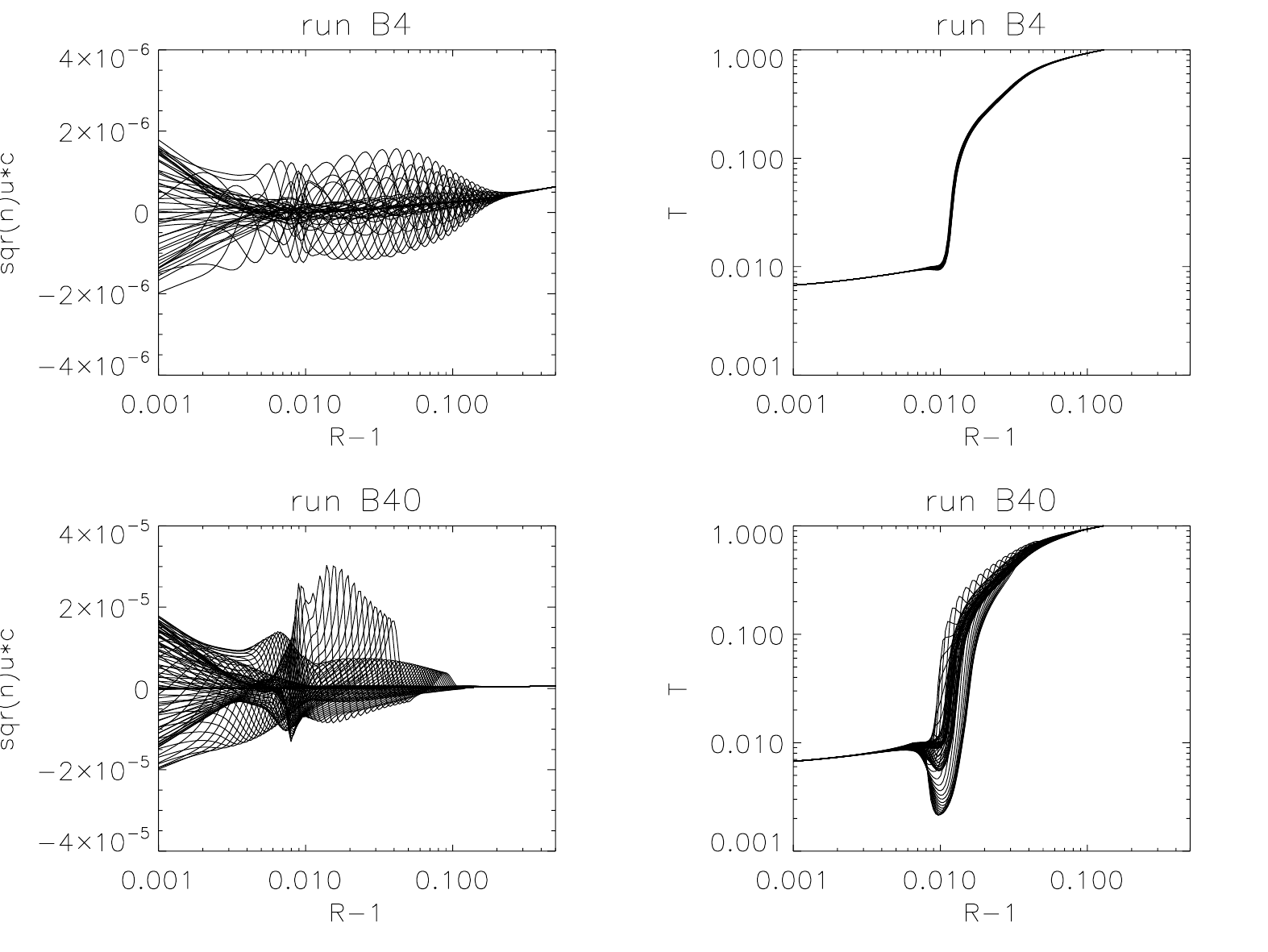}
\caption{Evanescent waves propagating through the full wind model.
Top: run B4, base amplitude 4 m/s (time interval (60,60.5)); bottom: run B40, base amplitude 40 m/s (time interval (60,60.25);
left: velocity profiles ponderated by density and sound speed $(\sqrt(n)u c_s)$;
right: temperature profiles. Note run B40 stops after time $t=60.3$ (see next figure).
}
\label{fig3}
\end{center}
\end{figure}

We consider as in Section 3.1 monochromatic waves: the velocity is given a sinusoidal pattern, after a transient of a half period. 
We deal only with the evanescent case, since it is the case which shows the largest difference with the atmosphere without heat sources.
The base period is fixed to be $\tau=10 \ \ min$.
Since the cut-off period below the TR is $\tau_0=6\ -\ 8\ \ min$, this implies that the wave is evanescent everywhere in the dense layers.
The base amplitudes are either 4m/s or 40m/s.
Top panels in Figure~\ref{fig3} show the low amplitude, the bottom panels the high amplitude. 

\begin{figure}[ht]
\begin{center}
	\includegraphics [width=\linewidth]{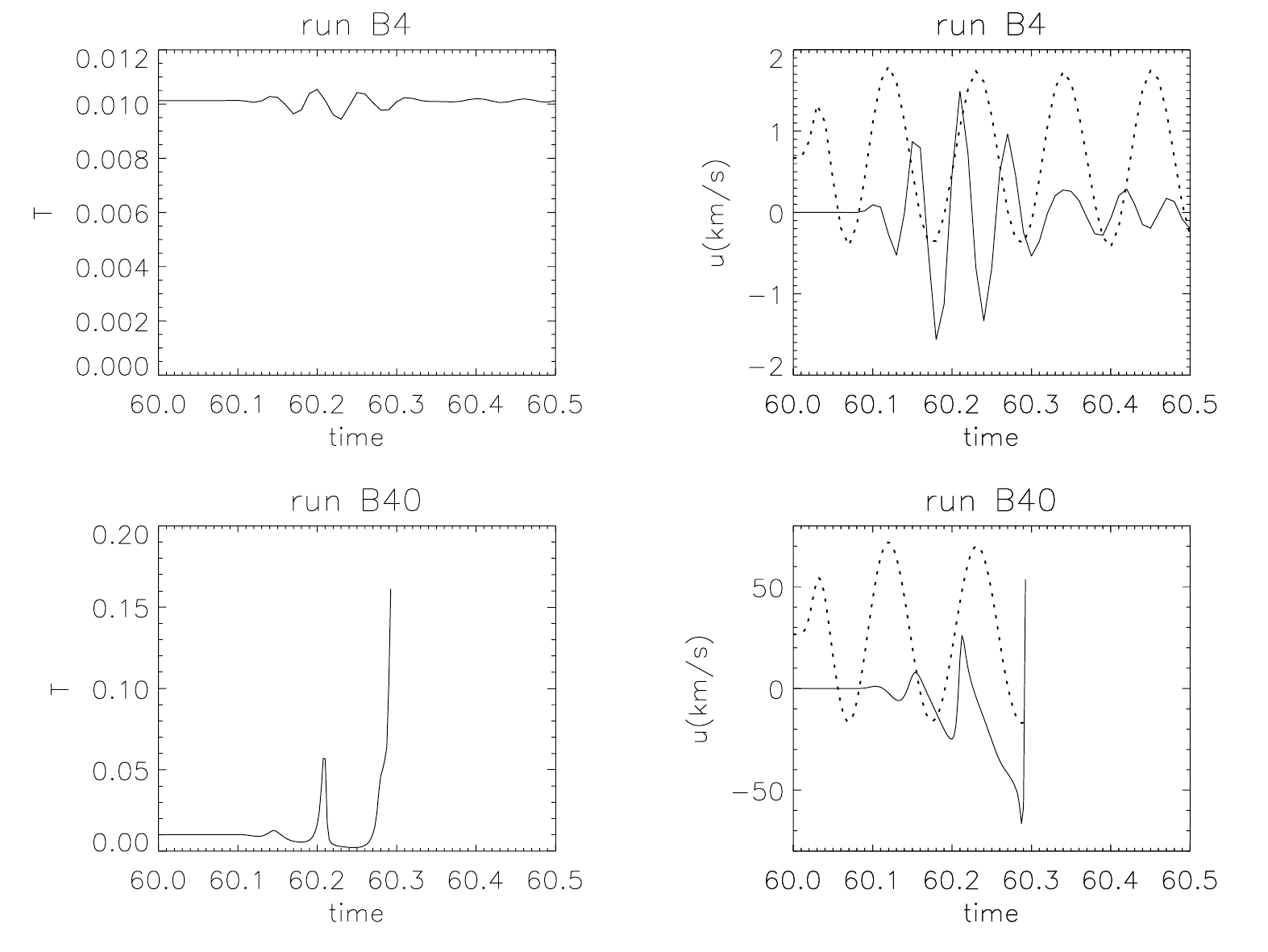}
\caption{Evanescent waves propagating through the full wind model.
Temperature and velocity at TR versus time;
dotted: base velocity;
top: run B4; bottom: run B40
}
\label{fig3a}
\end{center}
\end{figure}

In both runs, the TR adopts as in the relaxation model a period of $6 \ \ min.$, about the cut-off period (fig.~\ref{fig3a}. The dotted fluctuations (not at scale) show the base velocity fluctuations in both runs, for comparison.

Note that a peculiar behaviour of density and temperature is observed at the base: a transient systematic drift is superposed to the fluctuations. This drift modifies the mean gradients near the base, and allows the atmosphere to solve the contradiction imposed by the incoming characteristics which have in-phase fluctuations of velocity, temperature, density, while upward propagating eigenmodes have phase-lags between the different quantitites.

Run B4 shows two phases, one with about three large oscillations, while the rest of the fluctuations is of much smaller amplitude. Run B40 -- with the largest amplitude -- stops before the before second phase begins, because the profiles become too stiff for the numerical setup after the last temperature minimum, that is, after 60.25. (Although the calculation still proceeds up to time t=60.3, the time interval between 60.25 and 60.3 is unphysical, being invaded by numerical noise).
Note the prominent nonlinear features of the large amplitude run: sawtooth velocity pattern, and peaks for the temperatures maxima, large temperature troughs.

The amplitude being larger, one might suspect that the spatial scheme is not able to cope with too large and/or steep velocity/temperature/density profiles. 
However, a point deserves attention, namely the nonlinear relation between the oscillation amplitude at the TR and at the base: the strongest maximum of the velocity in run B4 is about 1.5 km/s, while the corresponding maximum is about 25 km/s, which is larger than the factor 10 which relates the base amplitudes.
Moreover, the beginning of the first phase 
is superficially similar to a (short) exponential growth in both runs.
So the hypothesis that the phenomenon we see is a true physical instability deserves to be examined in detail.

\subsection{Relaxation model again}

To see whether the result is generic or not, we replace now the energy equation by the plain temperature equation with a coupling to a given temperature profile, as already considered in section 3.1: in other words, we return to the relaxation model. 
However, we start now from run S3, with the temperature profile obtained through the full energy equation.
Hence, the difference with the runs just discussed in the previous subsection is that only the adiabatic terms will react to the impinging wave, since the energy terms are not present. But, again, the starting TR structure will be the same.

Recall that the extra coupling term in the temperature equation is of the form $-\kappa (T-T")$ (eq.~\ref{temperature}). 
We insist that $\kappa$ is a small parameter, which, in practice, gives this term a value which is never larger than $20\%$ of the physical conductivity, hence it should not modify substantially the growth rate of the instability, if it is there.

\begin{figure}[ht]
\begin{center}
	\includegraphics [width=\linewidth]{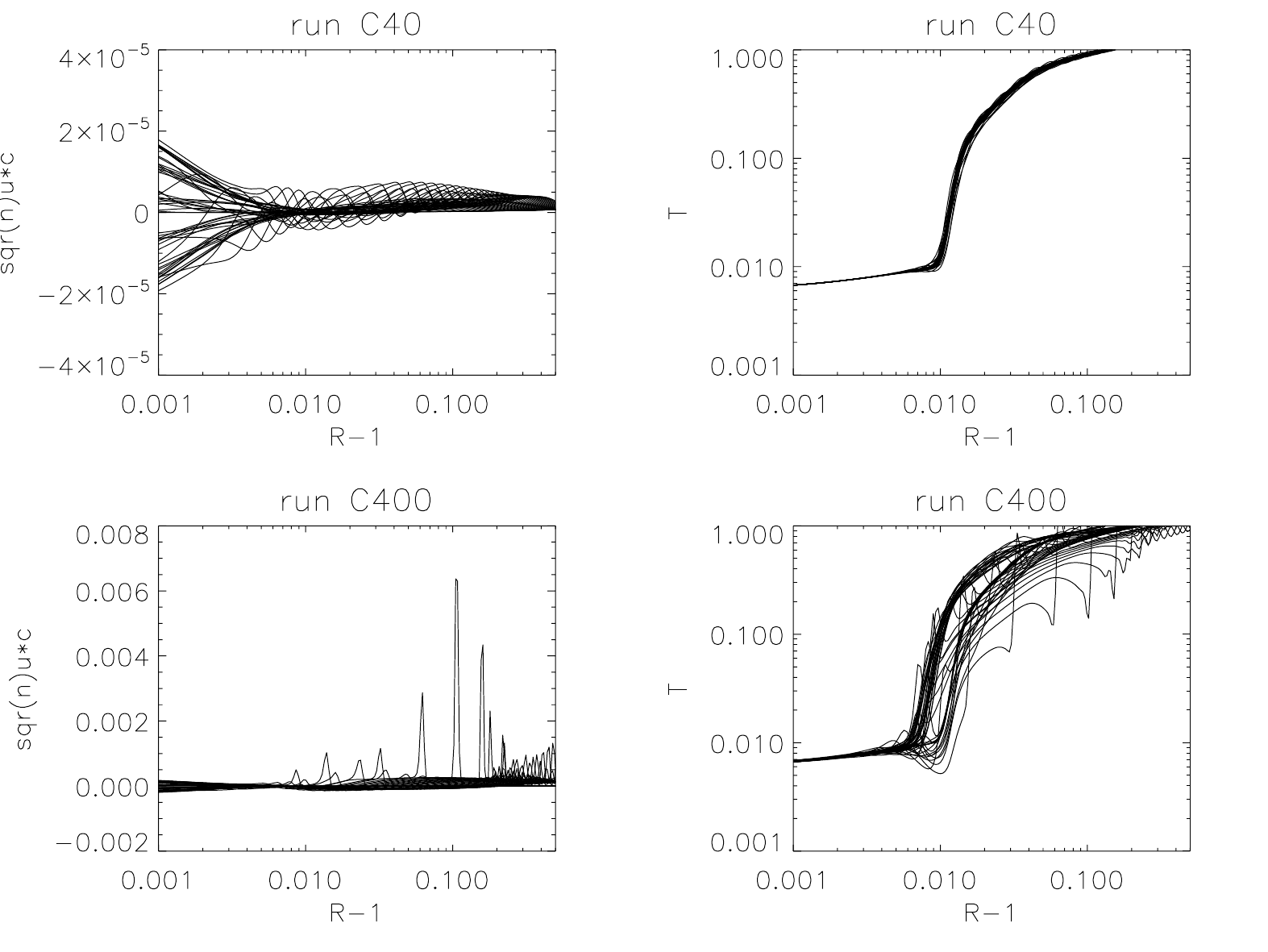}
\caption{Evanescent large-amplitude waves without energy terms, coupling with stationary temperature profile.
Top: run C40, base amplitude 40 m/s; bottom: run C400, base amplitude 400 m/s
left: velocity profiles ponderated by density and sound speed, $\sqrt(n)u c_s$;
right: temperature profiles. 
}
\label{fig4}
\end{center}
\end{figure}

The result, as shown in top of Figure~\ref{fig4}, shows that now the large amplitude wave (same as B40) arriving at the TR remain stable (run C40, top panels): the TR oscillations are actually much smaller than previously. 
Most importantly, when we increase the amplitude again by a factor 10 (run C400), we still obtain TR stability, with shocks propagating in the high corona (figure~\ref{fig4}, bottom panels).

\section{Discussion}

\begin{figure}[ht]
\begin{center}
	\includegraphics [width=\linewidth]{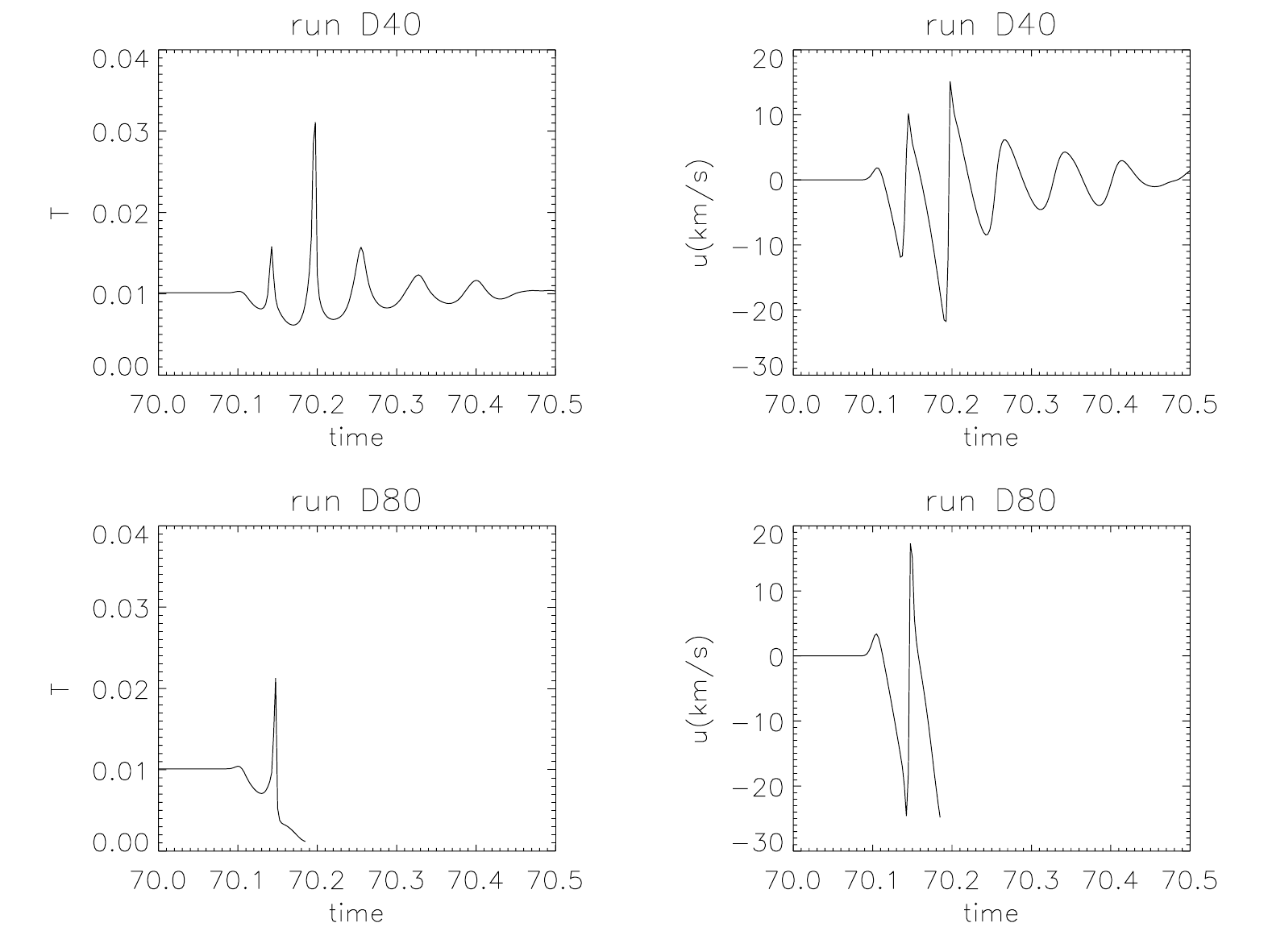}
\caption{Evanescent waves, full energy equation, double resolution;
temperature and velocity at TR versus time;
top: run D40, base amplitude 40m/s; bottom: run D80, base amplitude 80m/s
}
\label{fig5}
\end{center}
\end{figure}

\begin{figure}[ht]
\begin{center}
	\includegraphics [width=\linewidth]{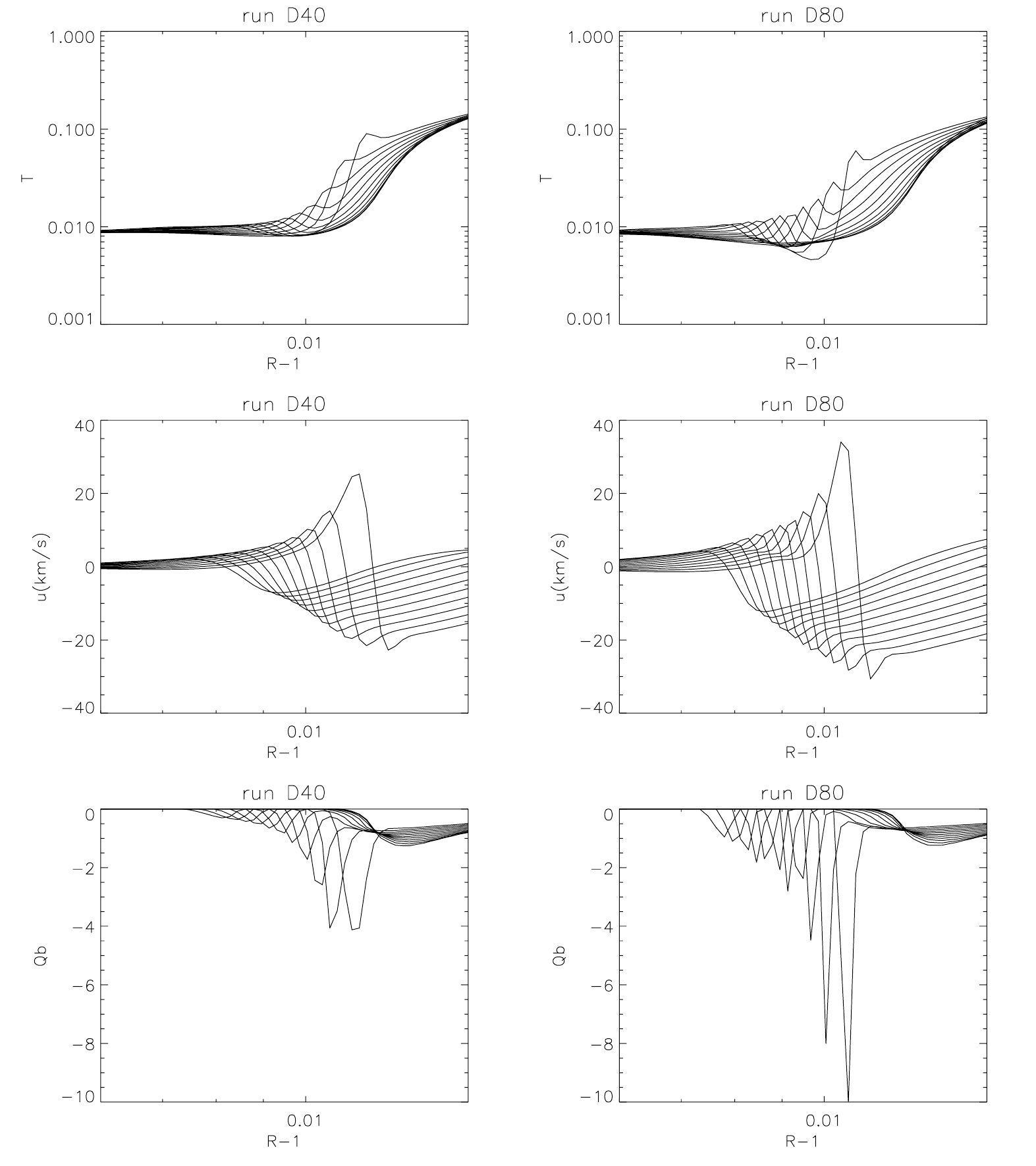}
\caption{Evanescent waves, full energy equation, double resolution;
successive profiles of temperature (top), velocity (middle) and cooling term as in eq. (3) and (6) (bottom),
in the vicinity of the TR during time interval 70.125 and 70.15;
left: run D40, base amplitude 40m/s; right: run D80, base amplitude 80m/s
}
\label{fig6}
\end{center}
\end{figure}

It is rather clear the response of the TR to impinging transient evanescent waves is very different, depending on whether extra-adiabatic terms are present or not.
The instability observed is nonlinear, that is, it shows up apparently only when waves have a sufficiently high amplitude.

To establish more firmly the physical status of the instability, we redo the experiment with large amplitude wave (B40), but now with double resolution, dividing each mesh in two parts. Asking for the same domain size, we thus end up with $N=599$ grid points instead of $N=300$.

The result is as shown in top of figure~\ref{fig5}: stable (run D40). We thus increase the base amplitude by a factor two (run D80), and now recover the instability of velocity and temperature (bottom figure). 
It is interesting to examine in detail the two successive phases in run D40 (the stable one), because they show all characteristics of run B40 (nonlinearities), but during a longer time, since it escapes numerical catastrophy.
During the first (nonlinear) phase, temperature and velocity are very different: temperature shows symmetric spikes (only for maxima), while velocity shows sawtooths.
During the second phase (which does'nt exist for run D80), one finds ordinary quasi-monochromatic oscillations of smaller amplitude for both fields.
The early velocity sawtooth pattern reveals shock formation, while the spiky appearance of the temperature pattern reveals the (nonlinear) advection of the temperature "wall" by the velocity field.
There is a last (crucial) feature shown by temperature, that is, the dangerous decrease of the temperature troughs, which disappears in the second phase for run D40, and finally leads to the end of run D80.

In fact, it seems that B40 is very close to run D40, the only difference being that D40 is able to pass the dangerous phase with the largest fluctuations, and not B40.
So, we can safely conclude that, while this difference is due to resolution change (and associated filtering change), the early nonlinear large response of the TR to upward propagating waves is physical, i.e., not due to the details of the spatial schema.
 
Note that the observed instability has an amplitude threshold, but that the response probably depends on the incoming perturbation being transient, as the spectrum is changing with time as time elapses after starting the injection (the signal is not purely monochromatic, beginning at a finite time).
Note also that the most striking difference between the low amplitude run (run B4) and the higher amplitude run (B40), as shown in figure~\ref{fig3} and also figure~\ref{fig3a} is the appearance of large cold troughs at the foot of the TR. 

While the temperature peaks (see figure~\ref{fig3a} or figure~\ref{fig5}) can be attributed to the (nonlinear) advection of the TR temperature "wall" by the large velocity fluctuations, what is the origin of the (growing, unstable) temperature troughs?
We propose that the cause both of the growth of the cold troughs is the radiative cooling term. This has the immediate advantage of explaining the nonlinearity of the response to the wave base amplitude. Indeed, the cooling term dependence on temperature is very nonlinear in the low temperature range (see section 2), with a cut-off at $10^4 K$ (eq~\ref{tstar}). 

Here is the detailed scenario we propose.
Consider the upward propagating wavefront resulting from a perturbation at the base, (for instance the beginning of our monochromatic injection) which is steepening progressively at it propagates, at a rate increasing with altitude (as its amplitude is growing with height, due to rough energy density conservation). 
We know that this progressive wave is not really equivalent to a plain sound wave, since it is a gravito-acoustic one, but we will take this into account later.

If the wave was submitted only to the usual adiabatic processes, then it would proceed as just said: propagating, growing, steepening. But there are extra terms in the energy equation. The conductive term is definitely negligible below the TR, as well as the mechanical flux term.
Note that the radiative cooling term remains strictly zero, but only as long as temperature remains below the threshold (here $10^4K$, eq.~\ref{tstar}).

Hence, this remains so all the way up to the TR, if the initial amplitude of the wave is small enough for the threshold $T\star = 10^4K$ not to be trespassed during propagation.
But, in the opposite case, then the wave crest is cooled immediately, with a rate proportional to the local density of the plasma.
Hence the temperature maxima of a large amplitude wave will be cooled both sooner (at lower altitude) and more rapidly than a small amplitude one.

\begin{figure}[ht]
\begin{center}
	\includegraphics [width=\linewidth]{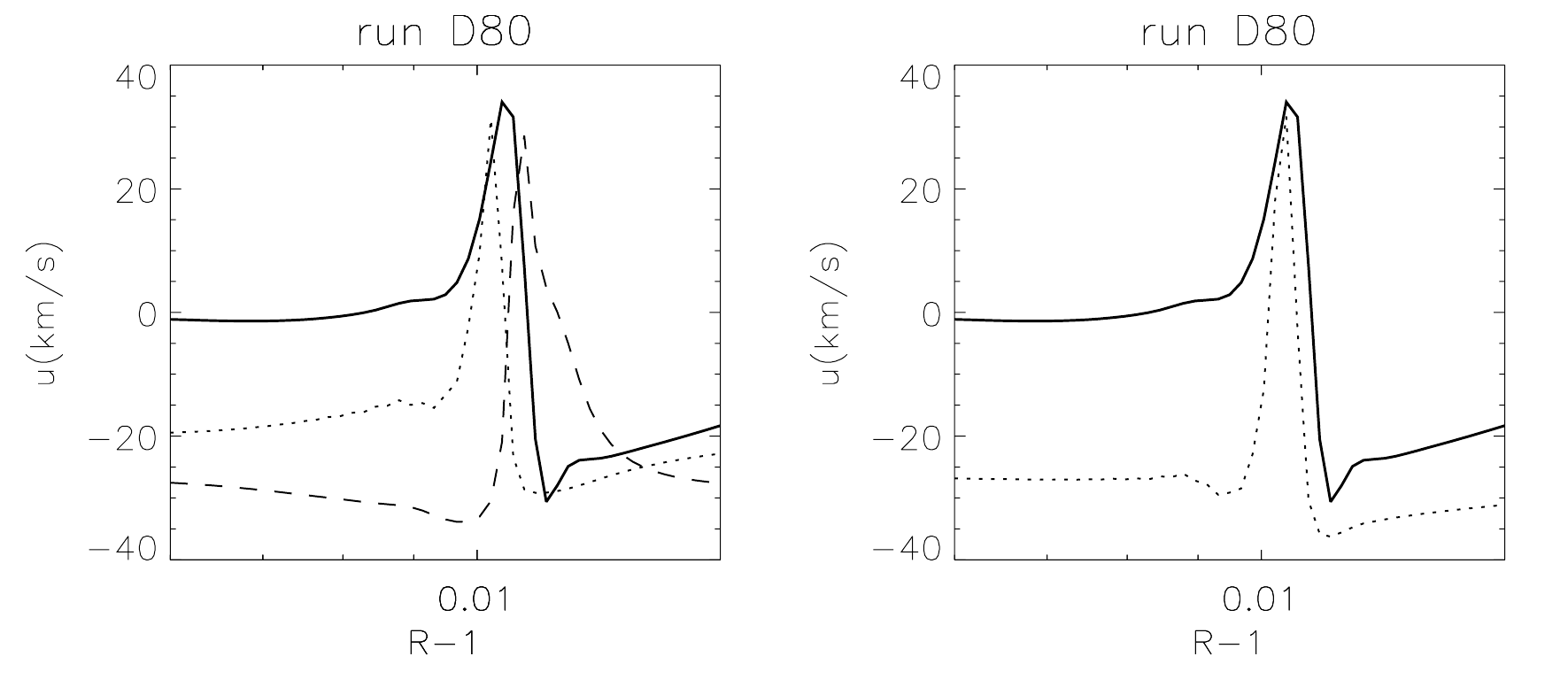}
\caption{run D80:
Instantaneous profiles of velocity, density, temperature and pressure at t=70.15;
left: velocity (thick line), density fluctuation (dotted), temperature fluctuation (dashed line); 
right: velocity (thick line), pressure fluctuation (dotted)
}
\label{fig7}
\end{center}
\end{figure}

What is the effect of the temperature drop when the wave crest is cooled down?
Figure~\ref{fig6} shows for runs D40 et D80 10 successive profiles at early times, from time t=70.125 to t=70.15, of temperature, velocity and cooling term in the vicinity of the TR.
The time interval corresponds to the first large oscillation at the TR, as seen in figure~\ref{fig5}.
It is seen that the growth of the cooling term and the growth of the temperature and velocity wavefront go together. 
Recall that both the conductive and mechanical heating terms are negligible in the region discussed, as stated above.


Now, consider for a moment that gravity can be neglected, i.e., that the wave frequency is high. Assume again local temperature is just below the threshold for radiative cooling. Also, consider as initial conditions not a progressive wave, but, instead, a local compressive movement (with no initial pressure fluctuation). This movement indeed will be accelerated as soon as the cooling threshold is trespassed, because the pressure increase which should resist the compressive movement is suddenly missing. 

But these conditions are not those of the progressive waves propagating from the base.
This is because (If gravity is neglected) then pressure, temperature, density and velocity fluctuations are in phase; in that case, cooling of the temperature maxima will affect the wave, but will not lead to an unstable compression.

What finally explains the instability in the present problem is that we are dealing not with a high-frequency wave, but with a wave at frequency close to the cut-off. 
In that case, density, pressure, velocity are no longer in phase, as may be seen in figure~\ref{fig7} which shows in details the different profiles at the last time of the series shown in figure~\ref{fig6}.
The left panel shows that the temperature crest is leading the velocity crest, while the density crest lies behind it. If we imagine now that the temperature crest is decreased by the radiative cooling, this will decrease the right foot of the pressure bump which should maintain the wave form. Hence the fluid ahead of the wave is accelerated toward the pressure drop, which is exactly what we see. This in turn raises the pressure, hence pushes again the temperature above the threshold, etc...

This argument explains also why injecting evanescent waves leads to instability, and not injecting propagative waves. This is because injecting evanescent waves generate much more waves with the cut-off frequency than the propagative waves (Fleck and Schmitz 1991, Kalkofen et al 1994) and because the instability occurring here requires that temperature, density and velocity be out of phase as found above in the transient propagative waves at the cut-off frequency.

Now, this instability clearly stops working when it enters the TR. 
This is because In the TR, conductive terms become important, and are able to balance  cooling, so the instability stops.

Hence, the instability has nothing to do with the TR itself: it is not an instability of the TR, but the instability of the (quasi-)isothermal, low temperature part of the atmosphere, which should occur whenever there is a layer which is optically thin, and cold enough, so that radiative cooling is active, and only it. 
Finally, the TR stops the instability, because of the stabilizing effect of thermal conduction there.

Also, in the cases (say, D40) where the evolution can be followed all the through the unstable phase, why does the process not repeat indefinitely for the next wavefronts arriving from the base? Why does the unstable phase stop after two or three periods? We think this last point has to do with the progressive disappearance of the cut-off part of the spectrum with time (Fleck and Schmitz (1991), Kalkofen et al. (1994)), that is, it is not, again, related to the TR itself.

The present scenario explains well our simulations, but what about the real sun?
In the real sun, there is a physical limit to the radiative cooling term which is used here, which has not only a temperature threshold as considered here, but also a density threshold, since in the densest layers the plasma becomes optically thick.

There are several ways to take this into account. The simplest is to introduce a dependence of the cooling term on density in the pressure equation.
A first possibility is as follows: 
$-n^2 \Lambda(T)$ becoming $n^2 n^\star/(n+n^\star)  \Lambda(T)$.
With this prescription, the cooling rate becomes density-independent when density is larger than $n^\star$. Note that if we set $n^\star$ close to the TR density itself (typically, $10^8 cm^{-3}$), its very structure will be strongly modified.
We have chosen  $n^\star = 10^{11} cm^{-3}$, and found that the growth rate of the first oscillations in a run equivalent do D80 was not basically changed.
A further case would be to replace $-n^2 \Lambda(T)$ by $(nn^\star/(n+n^\star)^2  \Lambda(T)$, which would completely switch off the term at layers denser than $n^\star$.

To summarize, the upward propagating wave packets generated by base perturbations will implode only in the layers which are comprized between the optically thick region and the TR where conduction becomes large, which stops the instability.
Hence the instability should be most observable when and where the TR is high enough above the optically thick layers.
This should provide an observational check of the present work.

Work is in progress to include magnetic effects, and generalize our study to an axisymmetric corona.

\begin{acknowledgements}
      Three of us (R.G., J.L., R.P.) thank warmly G. Belmont, S. Galtier for valuable comments at early stage of this work, S. Rifai Habbal for several stimulating discussions on observational data. One of us (RG) thanks S. Leygnac, and M. Velli for fruitful discussions on physics and numerics.
\end{acknowledgements}

\end{document}